\documentstyle[12pt,aaspp4]{article}
\begin{document}

\title{MICROLENSING OF BLENDED STELLAR IMAGES}

\author{Przemys\l aw Wo\'zniak and Bohdan Paczy\'nski}
\affil{Princeton University Observatory, Princeton, NJ 08544--1001, USA}
\affil{e-mail: wozniak@astro.princeton.edu, bp@astro.princeton.edu}

\begin{abstract}

The current modelling of single microlensing light curves neglects the
possibility that only a fraction of the light is due to the lensed star,
the remaining being due to a close, unresolved blend, which may
be related or unrelated to the lens.
	
Unfortunately, the effects of blending are significant
as all microlensing experiments choose very crowded fields as their targets.
In this paper we point out a strong degeneracy of the fitting procedure
which makes it practically impossible to detect the presence of a blend
by purely photometric means, except in a small part of the parameter space.
Some blends may be detected by astrometric means, but the majority
have to be corrected for statistically.  The luminosity function reaching 
well below the ground based detection limit (with the HST) would be very
helpful.  The statistics of binary stars in the target population is
also important and this could be determined with
the repeating microlensing events.

If no correction is made then the event time scales,
the lens masses, and the optical depth are all systematically underestimated.

\end{abstract}


\section{Introduction}

Almost all model fitting of single microlensing light curves to the
data is done with a 4-parameter curve which assumes that the stellar
image is not blended (cf. Paczy\'nski 1996, and references therein).
However, the first two double lenses: OGLE \#7 (Udalski et al. 1994b)
and DUO \#2 (Alard et al. 1995) were found to have apparent images
made of at least two objects: the lensed star and another unrelated star.
Also, at least one single event, OGLE \#5, was found to be strongly blended
(Mao 1995, Alard 1996b).  According to Alard (1996b) most DUO events are likely
to be blended, and fitting data with unblended light curves introduces
a systematic bias in the estimate of event time scales and optical depth.
Other effects of blending were considered by Nemiroff (1994), Di Stefano
\& Mao (1995), and by Buchalter \& Kamionkowski (1996).

Another issue is the lensing of very faint stars which become detectable
only while lensed (Nemiroff 1994).  Such objects are missed in searches
like DUO (Alard 1996a), EROS (Auburg et al. 1993), MACHO (Alcock et al. 
1993) and OGLE (Udalski et al. 1993), but they are detectable in the
searches which use the ``image subtraction'' technique, like AGAPE
(Bouquet et al. 1996) and COLUMBIA--VATT (Crotts \& Tomaney 1996,
Tomaney \& Crotts 1996, 1997).  An elaborate theory of this technique was
developed by Gould (1996).

The aim of this paper is to demonstrate some of the practical
consequences of blending, and the limitations imposed by unknown
blends on the determination of lens parameters.

\section{The Model}

Let us consider an idealized situation with a variability due to
microlensing superposed on arbitrary constant background of whatever
nature: the sky, or the crowded field made of many stars or nebulae.
We adopt an approximation according
to which in a given small aperture (it may be profiled as the PSF, the
Point Spread Function) there is a well measured level of brightness 
long before and long after the microlensing event:
$$
F_0 \pm \Delta _0 N_0^{-1/2}, \hskip 1.0cm N_0 \gg 1 ,  \eqno(1)
$$
where $ \Delta _0 $ refers to a standard deviation of a single measurement,
and the number of measurements $ N_0 $ is very large. 
Therefore, $ F_0 $ as well
as $ \Delta _0 $ are known very accurately.  However, we do not know 
what fraction of $ F_0 $ is due to the star which is microlensed
while it is in its normal, unlensed condition.

Let the flux from the lensed star be given as
$$
F_s = F_{s0} A(t) , \hskip 1.0cm F_{s0} = f_s F_0 < F_0 , \eqno(2)   
$$
where the magnification due to microlensing is
$$
A(t) = { u^2 +2 \over u \left( u^2 +4 \right) ^{1/2} } , \hskip 1.0cm
u^2(t) = u_{min}^2 + \left( { t - t_{max} \over t_0 } \right) ^2 ,
\eqno(3)
$$
where $ u_{min} $ is the impact parameter in units of Einstein ring radius,
$ t_{max} $ is the time at which maximum magnification is reached, and
$ t_0 $ is the time it takes the lens to move with respect to the source
by one Einstein ring radius (Paczy\'nski 1996, and references therein).

Given a set of photometric measurements: $ (F_k, t_k) , ~ i = 1, ~ 2, ~ 3, ~
..., ~ n $, we would like to determine four parameters:  $ F_{s0}, ~ u_{min}, ~
t_{max}, ~ t_0 $.  In principle the value of the fifth parameter:
$ F_0 $ has to be determined as well.  However, in practice a good
microlensing search is conducted over a time interval much longer than
the microlensing time scale $ t_0 $, so we may simplify our task by
adopting $ F_0 $ as known.  In a common fitting procedure it is assumed
that $ F_{s0} = F_0 $, and $ F_0 $ has to be determined from the same set
of measurements as the other three parameters: $ u_{min}, ~ t_{max}, ~ t_0 $.

We assume that the errors (standard deviations) of the measurements scale as
$$
\Delta _k = \left( { F_k \over F_0 } \right) ^{1/2} \Delta _0 ,  
\hskip 1.0cm  \left( { F_k \over F_0 } \right) = (1-f_s) +f_s A(t_k), 
\hskip 0.5cm f_s \equiv { F_{s0} \over F_0 } , \eqno(4)
$$
where $ \Delta _0 $ is the error of a single measurement at intensity
$ F _0 $, $ f_0 $ is the fractional intensity of the lensed star
far from the microlensing event, and the magnification $ A(t_k) $ is given 
with the eq. (3).  Our task is to find out how accurately the 
lens parameters can be determined.
Instead of a massive Monte--Carlo simulation we adopt another approach.
First, the values of $ F_0, ~ \Delta _0, ~  f_s, ~ u_{min}, ~
t_{max}, ~ t_0 $ are fixed.  The first two are assumed to be known to
the observer, while the latter four are to be determined.  The $ \chi ^2 $
of the fit between a model and a string of quasi-data points
$ (F_{data,k}, t_k) , ~ k = 1, ~ 2, ~ 3, ~ ..., ~ n $ is approximated with 
the sum:
$$
X^2 = \sum\limits_{k=1}^n 
{{ \left( F_{model, k} - F_{data, k} \right) ^2 + \Delta _k^2 \over 
\Delta _k ^2 }} ,
\eqno(5a)
$$
$$
F_{model, k} = \left( F_0 - F_{s0} \right) + F_{s, k} ,
\eqno(5b)
$$
where $ F_{model,k} $ are the intensities which follow from
the model we are fitting,
the quasi-observations $ F_{data, k} $ are given with the eq. (2), 
and the measurement errors $ \Delta _k $ are given with the eq. (4).
Our task is to determine the values of four parameters: $ \alpha _{1,2,3,4} 
= \left( f_s, ~ u_{min}, ~ t_{max}, ~ t_0 \right) $ by minimizing $ X^2 $.
Note that because of measurement errors the value
of $ X^2 $ can never be zero.
It is minimized for the correct choice of the four parameters,
when its value is $ n $.  We shall also determine
the confidence ranges of the four parameters for which the value of $ X^2 $ 
is within a chosen range of its minimum.
For simplicity we assume that $ n $ equally spaced measurements
cover the time interval $ t_{max} - 2 t_0 < t < t_{max} + 2 t_0 $.

For small errors we can expand $ X^2 $ around its minimum to obtain:
$$
\Delta X^2 \simeq 
\sum\limits_{j=1}^4 \sum\limits_{i=1}^4 D_{i,j} 
\Delta \alpha _i \Delta \alpha _j , 
\eqno(6)
$$
where $ \Delta \alpha _i = \alpha _{model, i} - \alpha _{data, i} ~
( i = 1, 2, 3, 4 ) $ is a vector of the parameter differences between
the fitting and the model values, and
$$ 
D_{ij} = {F_0 \over \Delta _0^2}
\sum\limits_{k=1}^{n}
{1 \over F_k} \left[{ \partial F_k \over {\partial \alpha _i}}
{ \partial F_k \over {\partial \alpha _j}}\right] ,
\eqno(7)
$$
at $ \alpha _i = \alpha _{data, i} , ~ (i=1,2,3,4) $.
For a quadratic expansion the confidence ellipsoid
it describes scales linearly with the measurement errors.
The errors with which all model parameters are determined scale linearly 
with the quantity $ a \equiv \Delta _0 n^{-1/2} $, for $ n \gg 1 $
and $ \Delta _0 \ll 1 $.

The best MACHO and OGLE microlensing events had $ n \approx 60 $
and $ \Delta _0 \approx 0.04 $, for the corresponding $ a \approx 0.005 $
(cf. Udalski et al. 1994a, Alcock et al. 1995).
A more typical values were 
$ (n, \Delta _0, a ) \approx ( 30, 0.1, 0.02) $.
However, with the introduction of effective follow-up
observations by the PLANET (Albrow et al. 1996) and GMAN (Pratt et al. 1996).
a substantial increase of $ n $, a reduction of $ \Delta _0 $, and
the corresponding reduction of the parameter $ a $ have
already been achieved in some cases, and farther improvement
is expected in the near future.  

In the following section
we investigate some problems with the determination of model
parameters imposed by the finite value of the $ a $ parameter.

\section{The Degeneracy}

There are two regions in the parameter space for which there is
a near degeneracy, i.e. with any realistic
accuracy of the measurements it is not possible to determine the unique
values of all model parameters.

The first troublesome case is that of a large impact parameter.
In the limit $ u_{min} \gg 1 $ the eqs. (2) and (3) may be transforemed
as follows:
$$
A \approx 1 + { 2 \over u^4 } =
{ 1 + {2 \over \left[ u_{min}^2 + (t/t_0)^2 \right] ^2 }} ,
\hskip 1.0cm {\rm for } \hskip 0.5cm u \gg 1 ,
\hskip 0.5cm t_{max} = 0 .
\eqno(8)
$$
Within this approximation we can write (cf. eqs. 2-4)
$$
{ F(t) \over F_0 } = ( 1-f_s ) + f_s \times \left( 1 + { 2 \over
\left[ u_{min}^2 + (t/t_0)^2 \right] ^2 } \right) =
1 + { 2 f_s \over
\left[ u_{min}^2 + (t/t_0)^2 \right] ^2 } .
\eqno(9)
$$
It is straightforward to verify that substituting the parameters
$$
f_{s,1} = f_s C^4 , \hskip 0.5cm
u_{min,1} = u_{min} C , \hskip 0.5cm
t_{0,1} = t_0 C^{-1} ,
\eqno(10)
$$
into the eq. (9) we recover the same formula for the $ F_s/F_0 $.
Therefore, if the set of parameters: $ \left( f_s, ~ u_{min} , ~ t_0 \right) $
is a solution then the set
$ \left( f_s C^4, ~ u_{min} C , ~ t_0 C^{-1} \right) $ is also a
solution, where $ C $ is an arbitrary positive constant which
satisfies $ f_s C^4 \leq 1 $.

Analytical considerations seem to imply that the trouble sets in
when the impact parameter is very large, $ u_{min} \gg 1 $.   In reality
the situation is much worse.
An example of the problem is shown in Fig. 1a, in which the two light curves
shown with a solid and a dashed line, respectively, are almost identical.
The truly troublesome aspect of this example is the value of the impact
parameter: $ u_{min} = 0.5 $.  This implies that in practice the degeneracy
covers a broad range of impact parameters, and a seemingly robust microlensing
event with the peak magnification in excess of 2 may have a blend which is
photometrically undetectable in the event's light curve.

Another troublesome case is when the minimum flux, i.e. the flux
measured far from the microlensing event, is dominated by a blend or
by any background.  This is 
a generic case in very crowded fields, like the bulge of M31,
in which microlensing events are to be detected by the ``image subtraction''
method (Bouquet et al. 1996, Crotts \& Tomaney 1996, Tomaney \& Crotts
1996, 1997).
As the lensed object contributes little to the total light at minimum,
the microlensing event can be detected only when the peak magnification
is very large, i.e. the impact parameter is very small.  In this case
we have $ F_{s0}/F_0 \ll 1 $, $ A_{max} \gg 1 $, $ u_{min} \ll 1 $,
and a significant change in the
brightness occurs only close to $ t_{max} $ when the projected distance 
between the source and the lens is very small, i.e. when $ u \ll 1 $.
We have
$$ 
A \approx {1 + {1 \over u}} \gg 1 , 
\hskip 1.0cm {\rm for} \hskip 0.5cm u \ll 1 , \eqno(11)
$$
and therefore
$$
{ F(t) \over F_0 } = ( 1 - f_s ) + f_s A \approx 
1 + { f_s \over u } = 1 + { f_s \over
\left[ u_{min}^2 + (t/t_0)^2 \right] ^{1/2} } ,
\hskip 0.5cm {\rm for} \hskip 0.3cm f_s \ll 1 , \hskip 0.3cm t_{max} = 0 .
\eqno(12)
$$
It is straightforward to verify that substituting the parameters
$$
f_{s,1} = f_s C , \hskip 0.5cm
u_{min,1} = u_{min} C , \hskip 0.5cm
t_{0,1} = t_0 C^{-1} ,
\eqno(13)
$$
into the eq. (12) we recover the same formula for the $ F_s/F_0 $.
Therefore, if the set of parameters: $ \left( f_s, ~ u_{min} , ~ t_0 \right) $
is a solution then the set
$ \left( f_s C , ~ u_{min} C , ~ t_0 C^{-1} \right) $ is also a
solution, where $ C $ is an arbitrary positive constant which
satisfies $ f_s C \leq 1 $.
An example of such a case is shown in Fig. 1b.

A more quantitative way to analyze the degeneracy is to plot the confidence
contours in the $ t_0 - u_{min} $ and in the $ f_s - u_{min} $ 
planes.  Two sets
of such plots are shown in Fig. 2; they correspond to the two cases presented
in Fig. 1.  It is apparent that the contours are highly elongated, i.e.
a certain combination of model parameters may be determined with a reasonable
accuracy, while the other combination is subject to a very large error.
The photometric accuracy parameter $ a \equiv \Delta _0/\sqrt{n}=0.007$
was assumed, which may be achieved for example when $\Delta _0=0.05$
and $ n=50 $.

The time of maximum magnification, $ t_{max} $, and the peak
brightness, $ F_{max} $ are not subject to any degeneracy in their 
determination as long as there are many points within FWHM of the 
light curve.  Unfortunately, these two parameters are of no particular
significance for any inferences from the microlensing observations.
Those parameters which are significant: the event time
scale $ t_0 $ and the impact parameter $ u_{min} $ tend to have
strongly correlated errors, as shown in Fig. 2.  It is a general
property of blended microlensing events that some combination of the
$ (t_0,u_{min}) $ values can be determined with a good accuracy, while
some other can be determined only poorly.  In both cases of degeneracy
as described with the eqs. (10) and (13) the product $ t_0 \times u_{min} $ 
can be measured well, but the ratio $ t_0 / u_{min} $ cannot be determined
from the observations, as it can be equal to any number.  Therefore,
we cannot determine the values of $ t_0 $ and $ u_{min} $ in these cases,
only the value of the product of the two parameters.  Note, that in
accordance with this analytical reasoning the confidence countours
as shown in the $ t_0 - u_{min} $ plane in Fig. 2 appear to be streached
along a hyperbolic shape.

There are two dimensionless parameters which determine the shape
of a blended microlening light curve: the impact parameter $ u_{min} $,
and the fraction of minimum light contributed by the lensed source,
$ f_s \equiv F_{s0}/F_0 $.  For a chosen value of the accuracy parameter 
$ a = 0.007 $
we calculated values of the standard deviations in the determination
of all four parameters: $ F_{s0}, ~ u_{min}, ~ t_0 , ~ t_{max} $ as
a function of $ F_{s0}/F_0 $ and $ u_{min} $, and they are
presented in Fig. 3.  The lines are labeled with the values 
of standard deviation.  For example, in the upper left corner
the error in the determination of the fraction of light in the
lensed object, $ \delta F_{s0}/F_{s0} $ is equal 1.0 (i.e. a huge error)
along the uppermost line, and the error is modest at 0.1 along the
lowest line.  If the accuracy parameter is smaller by a factor 10,
i.e. if $ a = 0.0007 $, then the labels along all lines would be
reduced by a factor 10, i.e. the accuracy of the determination
of $ F_{s0}/F_0 $ would be 0.1 along the uppermost line, and 0.01
along the lowest line.  A reduction of the $ a $ parameter by a factor 10
can be accomplished by reducing the photometric error $ \Delta _0 $
by a factor 10, or by increasing the number of photometric measurements
$ n $ by a factor 100.

An inspection of the Fig. 3 reveals that only the time of maximum
magnification $ t_{max} $ can be well determined over a significant
fraction of the parameter space.  The other three parameters:
$ F_{s0}, ~ u_{min}, ~ t_0 $ are virtually impossible to measure
above the lines with the label 1.0.  Note that our choice of the
accuracy parameter $ a $ was optimistic by the standards of current
microlensing seraches.

As the confidence contours are so elongated (cf. Fig. 2) it is interesting
to check if the best combination of the interesting parameters can be
measured with a significantly higher accuracy than possible for each
of the parameters separately.  This is shown in Fig. 4, where the right
two panels present the lines labelled with the corresponding values of a
standard deviation for the best linear combination of $ (u_{min}, F_{s0}/F_0) $
(upper right panel) and $ (u_{min}, t_0 ) $ (lower right panel).  Note,
lines in these panels are placed much higher than the corresponding lines
in Fig. 3, i.e. the errors are strongly reduced by choosing the best
combination of the parameter pairs, corresponding to the short axis
of the elliptical confidence contours in Fig. 2.  On the other hand, the 
opposite choice, corresponding to the long axis of the elliptical confidence 
contour, and presented in the two left hand panels in Fig. 4, reveal errors
even larger than those corresponding to each parameter separately, as
presented in Fig. 3.

Let us now ignore the presence of blends, and let us 
follow the standard procedure, assuming that $ F_{s0} = F_0 $.
With this assumption we determine the values of the event
time scale $ t_0' $ and the impact parameter $ u_{min}' $.  The
parameters so determined may be compared with their true values
$ t_0 $ and $ u_{min} $ in Fig. 5.  It is clear that by neglecting
the blend we always overestimate $ u_{min} $ and we underestimate $ t_0 $,
with the error increasing towards lower values of $ F_{s0} / F_0 $,
i.e. stronger blending, and lower values of impact parameter.

With the blending effects being so important we should check when the
presence of the blend can be established by means of photometry alone.
It was done so with the two double lenses (OGLE \#7, Udalski et al. 1994b;
DUO \#2, Alard et al. 1995) and with one single lens (OGLE \#5, S. Mao,
private communication 1995, Alard 1996b).  Unfortunately, this task is 
difficult for events caused by a single point mass, as shown in Fig. 6,
where the change in the value of $ X^2 $ is shown as a function of
two dimensionless parameters: $ F_{s0}/F_0 $ and $ u_{min} $.
This is the difference between the $ X^2 $ values of a blended event
as fitted with and without a blend.  If the change is large then
photometry alone can clearly demonstrate that the lensed source has a blend.
The parameter space above the thick
line $ \Delta X^2 = 2.8 $ is photometrically
degenerate: in this region it is not possible to detect the presence of 
a blend with the photometric accuracy parameter $ a = 0.007 $ at the
confidence level 90\%.

\section{Discussion}

The main result of this paper is somewhat discouraging: with the currently
typical photometric coverage and accuracy the light curve of a single
microlensing event cannot be used to determine the presence of a blend
unless the impact parameter is small, $ u_{min} < 0.3 $, and the blend
may be photometrically undetectable even if the impact parameter is
very small (cf. Fig. 6).  In the real world we do not know the precise
value of the measurement errors, and there is a significant contribution
to the errors which is non-gaussian (Udalski et al. 1994a).  Therefore,
the photometric border of blend detectability as indicated by the thick
line in Fig. 6 is optimistic, with the true border located at even lower
values of $ u_{min} $.
Yet, the blending must be very common as demonstrated
by the two double lensing events (OGLE \#7, Udalski et al. 1994a; DUO \#2,
Alard et al. 1995), and by the recent analysis of the DUO results 
(Alard 1996b).  Unless the blending is somehow corrected for, the 
microlensing time scales $ t_0 $ are underestimated, and the impact
parameters are overestimated (cf. Fig. 5).  This implies that the lens masses
and the optical depth are systematically underestimated.

There are various ways in which the blending may be detected or at least
statistically corrected for, as discussed by Alard (1996b).  In case of a
double lens, with caustic crossings indicated by the light curve, a blend
reduces the apparent magnification between caustic crossings below the
theoretical minimum value $ A_{min} = 3 $
(Witt \& Mao 1995) -- this was apparent in OGLE \#7 and
in DUO \#2.  In case of any event, double or single, a blend can be
detected through a correlation between the image centroid and the apparent
brightness, as first noticed in DUO \#2 (Alard et al. 1995).  This
is possible if the angular separation between the lensed object and 
the blend is not too small, presumably no less than half a pixel, or so.
If the blend has a different color than the lensed star then it may be
uncovered with multi band lightcurves of a microlensing event (Buchalter
\& Kamionkowski 1996).
If the luminosity function well below the detection threshold is known,
say from the HST imaging, then the blending can be statistically corrected
for by placing artificial stellar images on the CCD frames.  

The last approach has its limitations.  As long as we do not know the
distribution of separations and luminosity ratios of binary stars in
the target population we can only put artificial stars randomly on a
CCD frame.  But this procedure will underestimate the number of very close
binary pairs.  Fortunately, if the detected number of microlensing events is 
very large we can use the microlensing itself to determine the statistics
of binary stars, as they will produce repeating microlensing events 
(Di Stefano \& Mao 1996).

\acknowledgments{This work was supported with the NSF
grants AST--9313620 and AST--9530478.}

\newpage


\newpage
 
\begin{figure}[t]
\plotfiddle{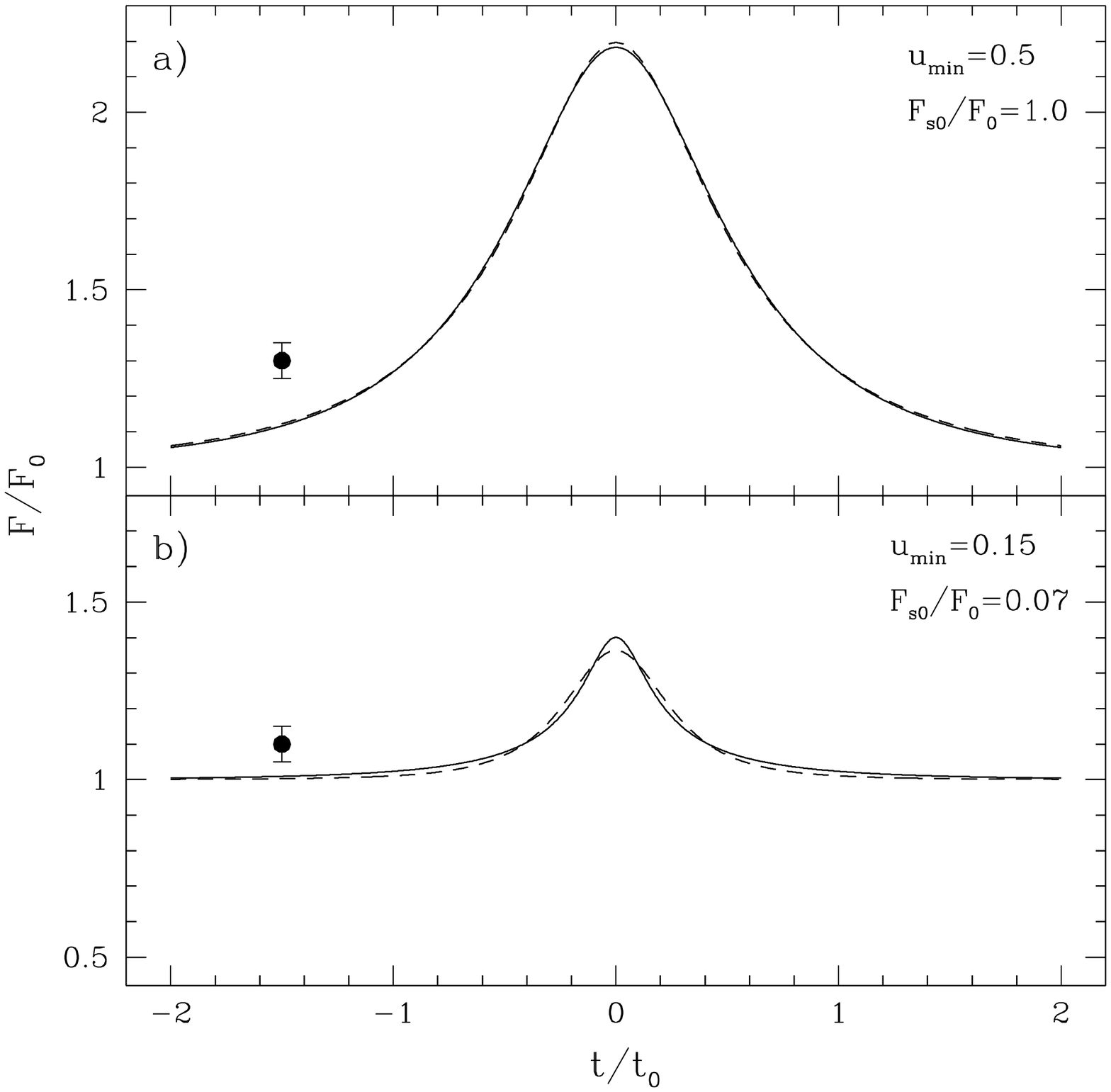}{8cm}{0}{50}{50}{-160}{-80}

\caption{Two examples of degeneracy are shown. (a) A large impact
parameter. The unblended event with $u_{min}=0.5$ and $t_0=1.0$
(solid) is well fitted by the model with $u_{min}^{\prime}=0.4$,
$t_0^{\prime}=1.15$ and $F_{s0}/F_0=0.73$ (dashed).  The difference
in the goodness of fit as defined with eqs. (5) is $\Delta X^2=0.52$. 
(b) A strong blending.  The solid line shows the
microlensing light curve with
the parameters: $u_{min}=0.15$, $t_0=1.0$,
$F_{s0}/F_0=0.07$.  The dashed light curve is the best
light curve for unblended event with
$u^{\prime}_{min}=0.97$, $t^{\prime}_0=0.30$.
The difference between the two corresponds to
$\Delta X^2=2.63$. In both plots a large dot with the error bar 
indicates $ \Delta _0 $, the photometric error at minimum light.
}

\label{fig:curves}

\end{figure}

\begin{figure}[t]
\plotfiddle{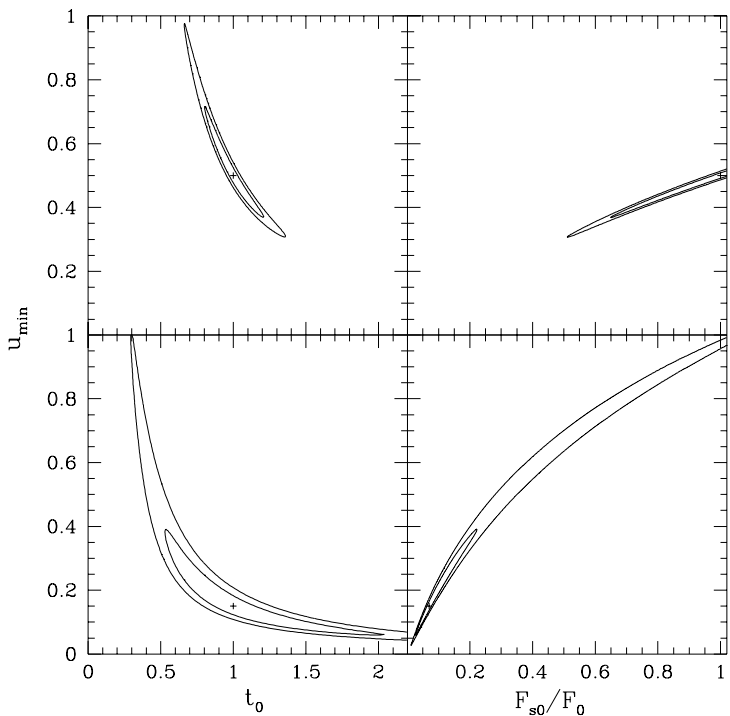}{8cm}{0}{50}{50}{-160}{-80}

\caption{Confidence contours in the $ (t_0 - u_{min} ) $ and 
$ (F_{s0}/F_0 - u_{min} ) $
planes for two cases of degeneracy from Fig.~1: a large impact parameter
($u_{min}=0.5$, upper) and a strong blending ($F_{s0}/F_0=0.07$, lower).
Confidence levels are 68\% and 90\% for one parameter of interest.
The photometric accuracy parameter $ a = \Delta _0 / \sqrt{n} = 0.007 $ 
was adopted.
}

\label{fig:cont}

\end{figure}

\begin{figure}[t]
\plotfiddle{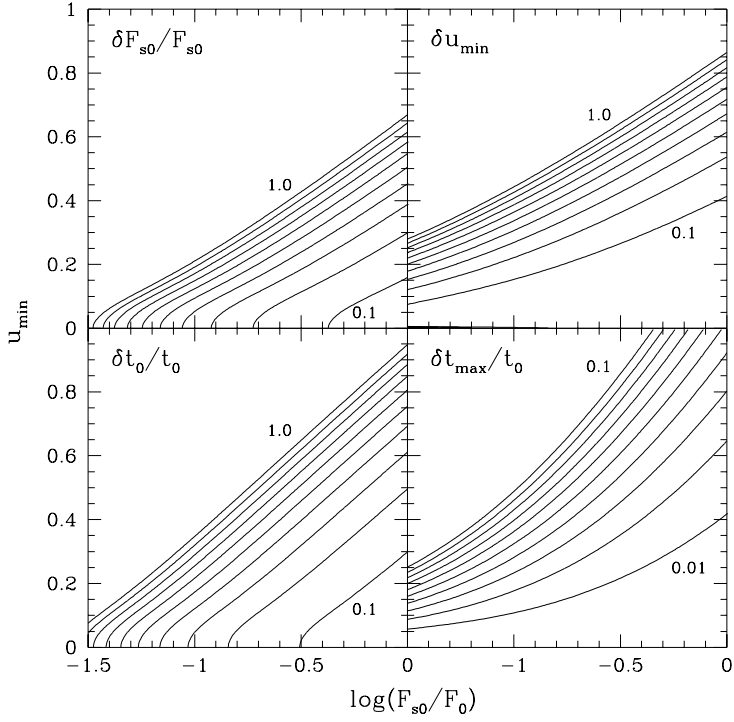}{8cm}{0}{50}{50}{-160}{-80}

\caption{The accuracy of the best fit parameters (68\% confidence)
as a function of $u_{min}$ and $F_{s0}/F_0$. The lines correspond to a
constant error in the best fit value determination for a given parameter.
For $F_{s0}/F_0$ and $t_0$ the fractional errors are shown.
The errors in $t_{max} $ are in units of $t_0$, while the errors in $u_{min}$ 
are in units of Einstein radius.
The lines are equally spaced in error value, between 0.1 to 1.0
for $F_{s0}$, $u_{min}$ and $t_0$, and from 0.01 to 0.1 for $t_{max}$.
All parameters are best measured when the impact parameter $ u_{min} $
and the blend contribution $ (F_0 - F_{s0})/F_0 $ are both small,
i.e. in the lower right hand corners of every panel.
The photometric accuracy parameter $ a \equiv \Delta _0/\sqrt{n} = 0.007 $
was adopted.
The ploted errors scale linearly with $a$ provided that the number of
photometric measurements $n$ is large and the photometric error $ \Delta _0 $
is small.
}

\label{fig:conf}

\end{figure}

\begin{figure}[t]
\plotfiddle{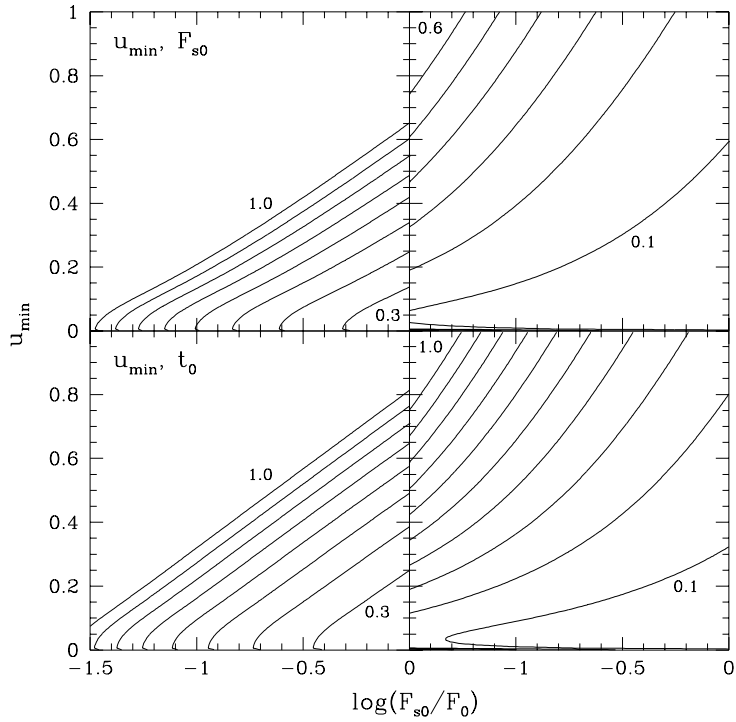}{8cm}{0}{50}{50}{-160}{-80}

\caption{The same as Fig.~3 but for two linear combinations of
$ (u_{min}, F_{s0}/F_0 ) $ (upper panels) and $ (u_{min}, t_0) $ (lower
panels), with the two combinations corresponding to the long axis
of confidence ellipse (the largest errors, left panels), and
the short axis of the ellipse (the smallest errors, right panels).
}

\label{fig:axes}

\end{figure}

\begin{figure}[t]
\plotfiddle{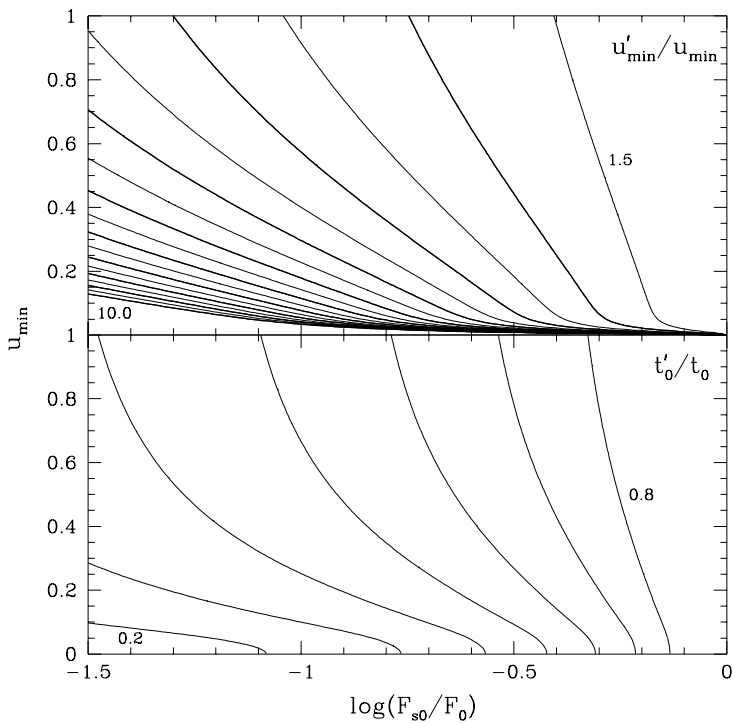}{8cm}{0}{50}{50}{-160}{-80}

\caption{The effects of fitting a standard light curve (without a
blend) to the blended
microlensing events. The plots show the ratio of the best fit value
of $u_{min}'$ (upper panel) and $t_0'$ (lower panel) as a function of 
$ \log (F_{s0}/F_0) $ and $u_{min}$.  If blending is ignored then
the impact parameter $ u_{min} $ is overestimated and the timescale
$ t_0 $ is underestimated.  The lines are labeled with the values
of $ u_{min}'/u_{min} $ and $ t_0'/t_0 $, and are equally spaced
in these values.
}

\label{fig:bias}

\end{figure}

\begin{figure}[t]
\plotfiddle{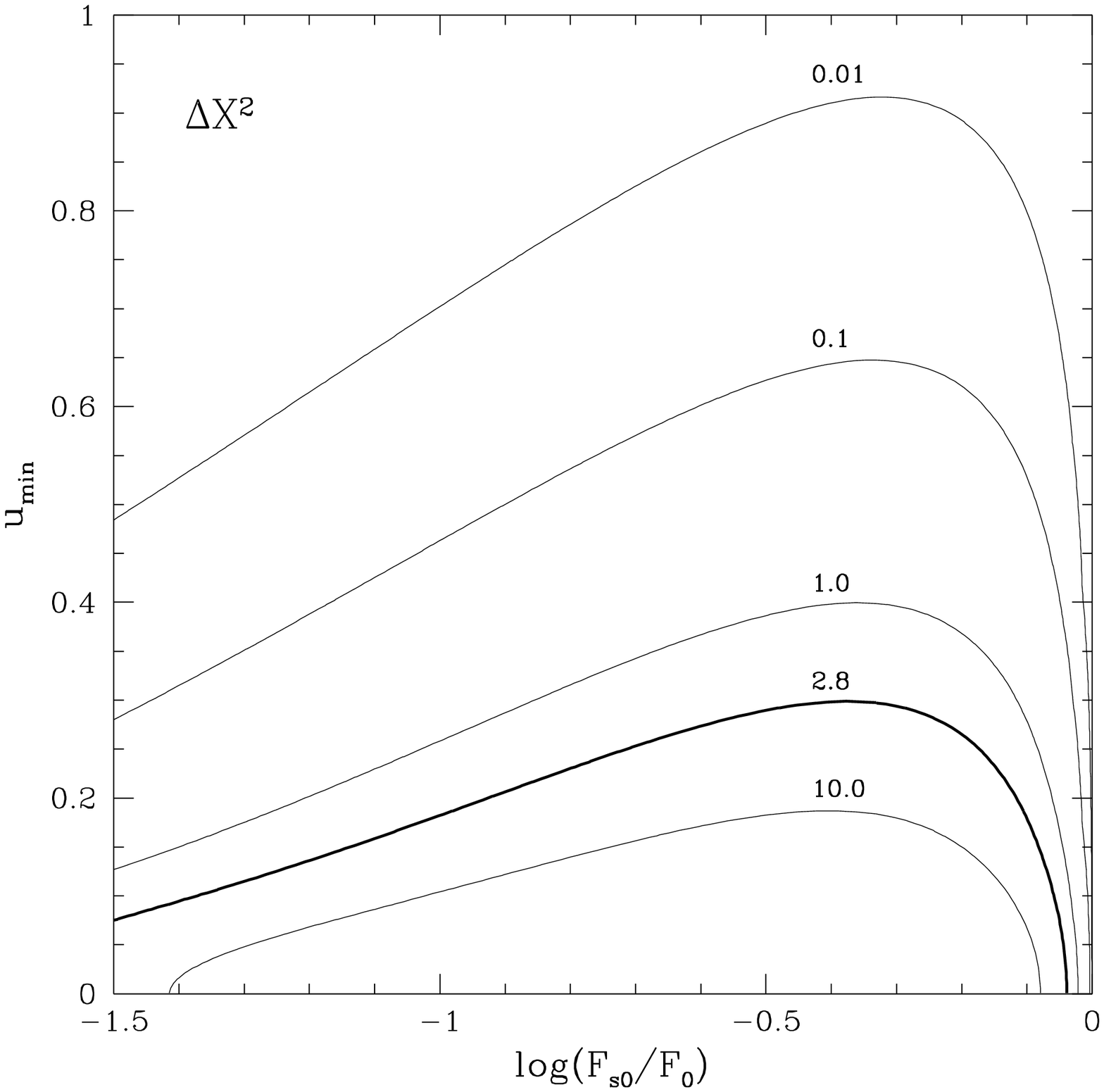}{8cm}{0}{50}{50}{-160}{-80}

\caption{
The difference $\Delta X^2$ between the best fit model with $F_{s0}=F_0$
and the best fit model with adjustable $F_{s0}$ is shown
as a function of the impact parameter
$u_{min}$ and the fraction of the lensed light $F_{s0}/F_0$.
Solid lines correspond to $\Delta X^2=0.01,0.1,1.0$, and 10.0 for
the photometric accuracy parameter $ a \equiv \Delta _0 / \sqrt{n} = 0.007 $.
It is impossible to discriminate statistically between the two types of 
models for $\Delta X^2 < 2.8$, i.e. above the thick line (90\% confidence).
The two areas of strong analytical degeneracy, one for large $u_{min}$ and the
other for small $F_{s0}/F_0$, are two parts of a single very large region.
When $F_{s0}/F_0$ approaches 1
the weak blend cannot be detected from a light curve alone,
but in this case the blend is of no practical consequence.
}

\label{fig:chisq}

\end{figure}


\begin{references}

\reference{} 

\reference{} 

\reference{} Alard, C. 1996a, 
in IAU Symp. 173: ``Astrophysical Applications
of Gravitational Lensing'', (Eds.: Kochanek, C. S., \& Hewitt, J. N.,
Kluwer Academic Publishers), p. 215
\reference{} Alard, C. 1996b, A\&A, in press (= astro-ph/9609165)     

\reference{} Alard, C., Mao, S., \& Guibert, J. 1995, A\&A, 300, L17  

\reference{} Albrow, M. et al. 1996, preprint: astro-ph/9610128, 
to appear in the Proceeedings of the 12th IAP Conference 
(Eds. R. Ferlet and J.-P. Maillard):
``Variable Stars and the Astrophysical Returns of Microlensing Surveys''

\reference{} Alcock, C. et al. 1993, Nature, 365, 621                 

\reference{} Alcock, C. et al. 1994, preprint: astro-ph/9512146               

\reference{} Aubourg, E. et al. 1993, Nature, 365, 623                

\reference{} Bouquet, A. et al. 1996,                                 
to appear in the Proceeedings of the 12th IAP Conference 
(Eds. R. Ferlet and J.-P. Maillard):
``Variable Stars and the Astrophysical Returns of Microlensing Surveys''

\reference{} Buchalter, A., \& Kamionkowski, M. 1996, ApJ, 469, 676

\reference{} Crotts, A., \& Tomaney, A. 1996, ApJ, 473, L87  

\reference{} Di Stefano, R., \& Mao, S. 1996, ApJ, 457, 93        

\reference{} Gould, A. 1996, ApJ, 470, 201   

\reference{} Nemiroff, R. J. 1994, ApJ, 435, 682          

\reference{} Paczy\'nski, B., 1996, ARA\&A, 34, 419

\reference{} Pratt, M. R. et al. 1996, 
in IAU Symp. 173: ``Astrophysical Applications
of Gravitational Lensing'', (Eds.: Kochanek, C. S., \& Hewitt, J. N.,
Kluwer Academic Publishers), p. 221 (= astro-ph/9508039)

\reference{} Udalski, A. et al. 1993, AcA, 43, 289                    

\reference{} Udalski, A. et al. 1994a, AcA, 44, 165                  

\reference{} Udalski, A. et al. 1994b, ApJ, 436, L103                  

\reference{} Tomaney, A., \& Crotts, A. 1996, AJ, 112, 2872 

\reference{} Tomaney, A., \& Crotts, A. 1997, 
to appear in the Proceeedings of the 12th IAP Conference 
(Eds. R. Ferlet and J.-P. Maillard):
``Variable Stars and the Astrophysical Returns of Microlensing Surveys''

\reference{} Witt, H. J., \& Mao, S. 1995, ApJ, 447, L105    

\end{references}
\end{document}